# Optimal Loss Reduction in Distribution Networks Using Conservation Voltage Reduction and Network Topology Reconfiguration

Rida Fatima, Hassan Zahid Butt, and Xingpeng Li

*University of Houston, 4222 Martin Luther King Blvd, Houston, 77204, Texas, USA*

**Abstract**

Conservation voltage reduction (CVR) and network topology reconfiguration (NTR) are widely employed to improve distribution system performance; however, existing approaches largely treat them independently, overlooking their coupled impact on load demand, voltage profiles, and power flow distribution, thereby limiting their overall effectiveness. This paper proposes a coordinated optimization framework for day-ahead operational planning of distribution networks, integrating CVR and NTR to enhance overall network efficiency and reduce active power losses in radial distribution networks. The problem is formulated as a mixed-integer conic programming model incorporating AC power flow constraints, voltage-dependent load representation, and radiality constraints. CVR is implemented to achieve load reduction through coordinated voltage control, while NTR redistributes line loading via optimal switching of controllable branches. The proposed framework is validated on the IEEE 33 and 123-bus distribution systems under varying load conditions. Results demonstrate that the coordinated approach consistently outperforms independent strategies, achieving up to 20.6% reduction in active power losses while maintaining voltage compliance and improving branch loading uniformity. These findings confirm that coordinated optimization provides an effective and scalable solution for enhancing efficiency in modern distribution networks.

**Nomenclature**

Sets:

| | |
|---|---|
| $S$ | Substations |
| $N$ | Nodes |
| $E$ | Branch |
| *E(i)* | Branch of which node *i* is the from-node |
| *E(j)* | Branch of which node *j* is the to-node |
| $T$ | Hours in day |
| $G$ | Generator |
| $PV$ | Solar PV |
| $B$ | Battery energy storage |

Indices:

| | |
|---|---|
| $s$ | Substation *s,* an element of set *S* |
| $n$ | Node *n*, an element of set *N* |
| $e$ | Branch *e*, an element of *E* |
| $t$ | Time *t*, an element of set *T* |
| $g$ | Generator *g*, an element of set *G* |
| $p$ | Solar *p*, an element of set *PV* |
| $b$ | Battery *b*, an element of set *B* |
| $k$ | Flexible branch *k,* an element of set *E* |

Parameters:

| | |
|---|---|
| $P_{n,t}^{D}$ | Active power load at bus *n* at time *t* |
| $Q_{n,t}^{D}$ | Reactive power load at bus *n* at time *t* |
| $r_{ij}$ | Resistance of branch *ij* |
| $x_{ij}$ | Reactance of branch *ij* |
| $soc_{min}$ | Minimum state of charge |
| $soc_{max}$ | Maximum state of charge |
| $V_{rated}$ | Rated voltage of substations |
| $V_{i,t}^{min}$ | Minimum voltage of substations |
| $V_{i,t}^{max}$ | Maximum voltage of substations |
| $I_{ij,t}^{max}$ | Maximum line current |
| $I_{ij,t}^{min}$ | Minimum line current |
| $C_{duration}$ | Battery Charging duration (h) |
| $D_{duration}$ | Battery Discharging duration (h) |
| $\eta_{b}^{chg}$ | Battery charging efficiency |
| $\eta_{b}^{dchg}$ | Battery discharging efficiency |
| $\vartheta$ | Conservation voltage reduction factor |
| $K_{p}^{Zp}$ | Active load constant-impedance fraction |
| $K_{p}^{IP}$ | Active load constant-current fraction |
| $K_{p}^{Pp}$ | Active load constant-power fraction |
| $K_{q}^{Zp}$ | Reactive load constant-impedance fraction |
| $K_{q}^{IP}$ | Reactive load constant-current fraction |
| $K_q^{Pp}$ | Reactive load constant-power fraction |
| $P_{g,t}^{max}$ | Maximum power produced by *g* at time *t* |
| $P_{g,t}^{min}$ | Minimum power produced by *g* at time *t* |
| $Q_{g,t}^{max}$ | Maximum reactive power produced by *g* |
| $Q_{g,t}^{min}$ | Minimum reactive power produced by *g* |
| $N_n$ | Number of nodes |
| $N_s$ | Number of substations |
| $P_{k,t}^{l}$ | Active ZIP load at bus $k$ and time $t$ |
| $Q_{k,t}^{l}$ | Reactive ZIP power load at bus $k$ and time $t$ |

Variables:

| | |
|---|---|
| $V_{n,t}$ | Voltage at node *n* at time *t* |
| $I_{ij,t}$ | Current on branch *ij* at time *t* |
| $j_{ij,t}$ | Status variable (binary) of flexible branch *ij* |
| $P_{s,t}^{Sub}$ | Power at substation *s* at time *t* |
| $Q_{s,t}^{Sub}$ | Reactive Power at substation *s* at time *t* |
| $P_{ij,t}$ | Power flow on branch *e* at time *t* |
| $Q_{ij,t}$ | Reactive power flow on branch *e* at time *t* |
| $P_{p,t}^{PV}$ | Power produced by solar *pv* at time *t* |
| $P_{b,t}^{discharge}$ | Active Power discharged by battery *b* at time *t* |
| $Q_{b,t}^{discharge}$ | Reactive power discharged by battery *b* at time *t* |
| $P_{k,t}^{inj}$ | Active power injections at node *k* at time *t* |
| $Q_{k,t}^{inj}$ | Reactive power injection at node *k* at time *t* |
| $P_{p,t}^{curt}$ | Curtailed solar power *p* at time *t* |
| $P_{b,t}^{charge}$ | Power used to charge the battery *b* at time *t* |
| $Q_{b,t}^{charge}$ | Charging reactive power of battery $b$ at time $t$ |
| $P_{g,t}^{gen}$ | Power produced by *g* at time *t* |
| $Q_{g,t}^{gen}$ | Reactive Power produced by *g* at time *t* |
| $c_{b,t}$ | Battery charging status *b* at time *t* (Binary) |
| $d_{b,t}$ | Battery discharging status *b* at time *t* (Binary) |
| $E_{cap}^{b}$ | Battery capacity in MWh |
| $E_{ene,t}^{b}$ | Battery energy at time *t* (MWh) |
| $E_{ene,init}^{b}$ | Initial energy stored in battery (MWh) |

## 1. Introduction

Distribution networks (DNs) are large-scale and complex systems responsible for delivering electrical energy from high-voltage transmission grids to industrial, commercial, and residential consumers. These networks are undergoing rapid transformation due to rising demand levels, extended feeder lengths, and the increasing penetration of distributed energy resources (DERs), particularly at medium- and low-voltage levels. Among these resources, photovoltaic generation introduces significant variability in power injections, with fluctuations occurring over short and long-time intervals [1]. This variability intensifies voltage regulation and power flow management challenges in modern distribution systems.

High penetration of DERs has been widely reported to exacerbate voltage-related issues in distribution networks, including overvoltage conditions [2], excessive operations of on-load tap changers [3], voltage instability in weak feeders [4], and accelerated degradation of voltage regulation equipment [5]. To mitigate these challenges, voltage control strategies in distribution networks are required and they are commonly categorized into decentralized and coordinated approaches. Decentralized methods rely on local measurements and autonomous control of legacy devices such as tap changers and switched capacitor banks, or predefined inverter-based control characteristics [6]. However, these approaches often overburden specific resources and lack system-wide optimality [7]. Coordinated strategies [8], in contrast, employ centralized, rule-based, or optimization-based frameworks to regulate voltage and power flows across the network [9]. Despite their advantages, planning-based coordinated schemes face limitations in adapting to uncertainty and variability introduced by high levels of renewable integration [10].

To improve operational efficiency in distribution networks, optimization-based techniques such as network topology reconfiguration (NTR) and conservation voltage reduction (CVR) have been extensively studied. Early work demonstrated that optimal network reconfiguration can significantly reduce power losses and improve load balancing through feeder switching actions [11], while more recent studies have leveraged mixed-integer and convex optimization formulations to achieve minimum-loss reconfiguration under radiality constraints [12], [13]. Metaheuristic and multi-objective approaches have also been applied to distribution system reconfiguration to simultaneously address loss minimization and reliability enhancement [14]. To handle practical system scale challenges, parallel computing frameworks have been introduced for reconfiguration under time-varying load conditions [15], while uncertainty-aware formulations have been proposed to improve system reliability in the presence of load variability [16].

More recently, mixed-integer second-order cone programming models have enabled the coordinated optimization of reactive power control and network reconfiguration in active DNs [17]. In addition, the role of inverter-based resources in supporting voltage regulation through reactive power control has been investigated [18], and linearized power flow and optimal power flow models have been developed to improve computational efficiency for active DN applications [19]. Dynamic reconfiguration strategies have further been explored to adapt network topology in response to time-varying load [20]. In parallel with NTR, CVR has emerged as a complementary technique that leverages voltage-dependent load behavior to achieve energy savings and loss reduction while maintaining acceptable voltage limits. At the transmission level, prior studies have further demonstrated that NTR can serve as an effective congestion management mechanism, reducing renewable energy curtailment and improving energy storage utilization under security-constrained operating conditions [21]. CVR, on the other hand, reducing feeder voltage levels within regulatory limits [22] can decrease peak demand and achieve long-term energy savings [23]. Traditional CVR implementations rely on rule-based voltage control using on-load tap changers, capacitor banks, and step voltage regulators operating on relatively slow time scales [24], [25]. While effective under conventional loading conditions, these approaches become increasingly challenged in networks with high DER penetration.

Recent studies have examined the interaction between DERs and CVR [26], demonstrating that inverter-based resources can either exacerbate voltage regulation challenges [27] or provide reactive power support to enhance CVR effectiveness [28]. These findings highlight the importance of coordinated optimization strategies capable of simultaneously accounting for voltage-dependent load behavior, DER operation, and network constraints. Moreover, the effectiveness of CVR is strongly influenced by load voltage sensitivity, which is more accurately represented using impedance-current-power (ZIP) load models rather than constant power assumptions.

While CVR and NTR have been widely studied as independent operational strategies, their coordinated application remains limited in existing literature. In practice, these mechanisms are inherently coupled through network physics. CVR modifies nodal voltages, which alters load demand and current magnitudes, thereby influencing power flow distribution and the optimal feeder configuration. Conversely, topology reconfiguration changes network connectivity and branch impedances, affecting voltage profiles and consequently the effectiveness of CVR. Motivated by this interaction, this work develops a unified optimization framework for day-ahead operational planning of distribution networks, enabling the joint adjustment of voltage levels and feeder topology.

The proposed framework is formulated as a mixed-integer conic programming model incorporating AC power flow equations and radiality constraints. Within this formulation, CVR is implemented through coordinated voltage control to reduce load demand, while NTR is realized via binary switching decisions that redistribute line loading across the network. By jointly optimizing these mechanisms over the scheduling horizon, the proposed approach captures the interaction between demand variation and power flow redistribution, resulting in improved loss reduction, enhanced voltage compliance, and more uniform branch loading compared to conventional independent implementations.

## 2. Network modeling & optimization formulation

The proposed optimization framework aims to minimize active power losses in a radial DN while incorporating network reconfiguration, CVR, voltage-dependent load behavior, and DER operation. The formulation is based on an AC optimal power flow representation and is relaxed into a mixed-integer conic programming structure to ensure computational tractability.

### *2.1 Objective Function*

To facilitate a conic-compatible formulation, auxiliary variables are introduced to represent squared voltage and current magnitudes as follows:

$$I^s_{ij,t} = I^2_{ij,t},\ V^s_{i,t} = V^2_{i,t}$$

This transformation eliminates square-root and bilinear nonlinearities and enables the branch-flow equations to be expressed in a form suitable for efficient solution. The optimization objective minimizes total active power losses across all distribution lines over the scheduling horizon and is expressed as (1).

$$min \sum_{ij\in E} \sum_{t\in T} r_{ij} I^s_{ij,t} \tag{1}$$

This function computes resistive losses by summing the product of branch resistance and the squared current magnitude on each branch. The quadratic current-loss relationship accurately captures physical line losses and provides a direct mechanism for loss minimization through voltage regulation, topology reconfiguration, and current redistribution.

### *2.2 Power Flow & Network Constraints*

The active and reactive power balance at each node is enforced through (2) and (3).

$$\sum_{(k,i)\in E}(P_{ki,t}-r_{ki}I_{ki,t}^{s})-\sum_{(i,j)\in E}P_{ij,t}+P_{i,t}^{inj}=0\ \forall i\in N,t\in T \tag{2}$$

$$\sum_{(k,i)\in E}(Q_{ki,t}-x_{ki}I_{ki,t}^{s})-\sum_{(i,j)\in E}Q_{ij,t}+Q_{i,t}^{inj}=0\ \forall i\in N,t\in T \tag{3}$$

Equation (2) ensures that the net active power injection at each node equals the difference between total incoming and outgoing active power flows, while (3) enforces the corresponding reactive power balance. Together, these constraints guarantee energy conservation at every node and time step. The voltage drop along each distribution line is modeled using

$$V_{i,t}^{s}=V_{j,t}^{s}+2\left(r_{ij}P_{ij,t}+x_{ij}Q_{ij,t}\right)-\left(r_{ij}^{2}+x_{ij}^{2}\right)I_{ij,t}^{s}\ \forall(i,j)\in E \tag{4}$$

which relates the squared voltage magnitudes at the sending and receiving nodes to the active and reactive power flows and the electrical parameters of the branch. Squared current and voltage magnitude variables, defined previously, are used to express the branch-flow equations in a conic-compatible form. These auxiliary variables enable direct modeling of line losses and voltage drops without introducing square-root nonlinearities. As a result, the voltage-power-current relationships can be formulated using quadratic and conic constraints, improving numerical stability and computational efficiency while preserving the physical characteristics of the distribution network. The relationship between branch current magnitude and power flow is captured by

$$P_{ij,t}^{2}+Q_{ij,t}^{2}\leq I_{ij,t}^{s}V_{i,t}^{s}\qquad\forall\ i,j\in E \tag{5}$$

ensuring consistency between current variables and active/reactive power transfer. These constraints collectively describe the AC power flow behavior within the DN under the adopted conic relaxation.

### *2.3 Nodal Power Injection Modeling*

The total active and reactive power injected at each node are defined in the following two equations respectively.

$$P_{k,t}^{inj}=\sum_{s\in S(n)}P_{s,t}^{Sub}+\sum_{p\in PV(n)}P_{p,t}^{PV}-\sum_{p\in PV(n)}P_{p,t}^{curt}-\sum_{b\in B(n)}P_{b,t}^{charge}+\sum_{b\in B(n)}P_{b,t}^{discharge}+\sum_{g\in G(n)}P_{g,t}^{gen}-P_{k,t}^{l} \tag{6}$$

$$Q_{k,t}^{inj}=\sum_{s\in S(n)}Q_{s,t}^{Sub}+\sum_{b\in B(n)}Q_{b,t}^{discharge}-\sum_{b\in B(n)}Q_{b,t}^{charge}+\sum_{g\in G(n)}Q_{g,t}^{gen}-Q_{k,t}^{l} \tag{7}$$

Equation (6) aggregates active power contributions from the substation, distributed generators, photovoltaic units, and energy storage systems, while accounting for curtailed photovoltaic power. Equation (7) similarly represents the reactive power injection from controllable sources. These expressions allow coordinated scheduling of multiple resources while maintaining nodal power balance.

*2.4 Voltage-Dependent Load Modeling*

To accurately capture the impact of voltage variations on demand, nodal loads are modeled using the ZIP formulation:

$$P_{k,t}^{l} = P_{n,t}^{D} * (K_{p}^{Zp} * \left(\frac{V_{k,t}}{V_{rated}}\right)^{2} + K_{p}^{Ip} * \left(\frac{V_{k,t}}{V_{rated}}\right) + K_{p}^{Pp}) \tag{8}$$

$$Q_{k,t}^{l} = Q_{n,t}^{D} * (K_{q}^{Zp} * \left(\frac{V_{k,t}}{V_{rated}}\right)^{2} + K_{q}^{Ip} * \left(\frac{V_{k,t}}{V_{rated}}\right) + K_{q}^{Pp}) \tag{9}$$

In this representation, active and reactive power demands are expressed as functions of nodal voltage magnitude normalized to the rated voltage. The quadratic, linear, and constant terms correspond to constant-impedance, constant-current, and constant-power load components, respectively. This modeling approach is essential for evaluating the effectiveness of CVR, as it directly links voltage regulation to demand variation and loss behavior.

*2.5 Operational and Security Constraints*

Safe and reliable network operation is ensured through the following constraints:

$$\left(V_{i,t}^{min}\right)^{2} \leq V_{i,t}^{s} \leq \left(V_{i,t}^{max}\right)^{2} \tag{10}$$

$$0 \leq I_{ij,t}^{s} \leq \left(I_{ji,t}^{max}\right)^{2} \tag{11}$$

$$P_{g,t}^{min} \leq P_{g,t}^{gen} \leq P_{g,t}^{max} \tag{12}$$

$$Q_{g,t}^{min} \leq Q_{g,t}^{gen} \leq Q_{g,t}^{max} \tag{13}$$

Equation (10) enforces nodal voltage limits, preventing overvoltage and undervoltage conditions. Equation (11) restricts branch currents to their thermal limits, ensuring conductor safety. Equations (12) and (13) define the minimum and maximum active and reactive power outputs of dispatchable generators, respectively, reflecting their physical operating capabilities.

*2.6 Photovoltaic Curtailment Constraints*

Photovoltaic power curtailment is constrained by

$$0 \leq P_{p,t}^{curt} \leq P_{p,t}^{PV} \tag{14}$$

ensuring that curtailed power remains non-negative and does not exceed the available photovoltaic generation at each time step. This formulation guarantees physically feasible curtailment decisions while allowing flexibility in network operation.

*2.7 Energy Storage System Modeling*

The operational behavior of energy storage systems is governed by the following constraints.

$$soc_{min} * E_{b_{cap}} \leq E_{b_{ene_t}} \leq soc_{max} * E_{b_{cap}} \tag{15}$$

$$c_{b,t} + d_{b,t} \leq 1 \tag{16}$$

$$0 \leq P_{b_{chg_t}} \leq \frac{E_{b_{cap}}}{C_{\text{duration}}} * ct \tag{17}$$

$$0 \leq P_{b_{dchg_t}} \leq \frac{E_{b_{cap}}}{D_{\text{duration}}} * dt \tag{18}$$

$$E_{b_{ene_{t=1}}} = E_{b_{ene_{ini}}} + (\eta_b^{chg} * P_{b_{chg_{t=1}}}) - P_{b_{dchg_{t=1}}} / \eta_b^{dchg} \tag{19}$$

$$E_{b_{ene_t}} = E_{b_{ene_{t-1}}} + \left(\eta_b^{chg} * P_{b_{chg_t}}\right) - P_{b_{dchg_t}} / \eta_b^{dchg} \quad t \geq 2 \tag{20}$$

$$E_{ene,T}^{b} = E_{ene,init}^{b} \tag{21}$$

Equation (15) constrains the state of charge within allowable limits. Equation (16) enforces mutually exclusive charging and discharging operation. Equations (17)-(18) limit charging and discharging power based on storage capacity and operating duration. The energy balance equations (19)-(20) update the stored energy across time while accounting for efficiency losses. Equation (21) ensures that each battery ends the scheduling horizon with the same energy it started with, preventing artificial energy gain or loss. Together, these constraints ensure realistic and physically consistent energy storage operation.

*2.8 Network Topology Reconfiguration*

In a topology reconfigurable network, certain lines can be opened or closed to optimize losses and improve reliability, while maintaining a radial network structure. This is represented in (22), where the branch current $I_{ij,t}^{s}$ is limited by the product of its maximum allowable current and a binary switching variable $j_{ij,t}$, which indicates whether the branch is active (1) or open (0).

$$0 \leq I_{ij,t}^{s} \leq \left(I_{ji,t}^{max}\right)^2 * j_{ij,t} \tag{22}$$

$$\sum_{i,j\in E} j_{ij,t} = Nn - Ns \tag{23}$$

Radiality of the network is enforced by (23). These constraints ensure that the resulting topology remains loop-free while allowing optimal switching actions. This formulation enables loss minimization through structural optimization without compromising network operability.

### *2.9 Conservation Voltage Reduction*

CVR is incorporated into the optimization framework by tightening the allowable lower voltage bound at each node, as expressed in

$$(1-\vartheta)\left(V_{i,t}^{min}\right)^2 \le V_{i,t}^{s} \le \left(V_{i,t}^{max}\right)^2 \tag{24}$$

The voltage reduction factor $\vartheta$ scales the minimum allowable voltage bound, thereby tightening the operational voltage range. This study considers a slightly relaxed lower bound to evaluate the trade-off between voltage regulation, load reduction, and loss minimization. This formulation enables controlled voltage reduction across the feeder while maintaining system feasibility. By driving nodal voltages toward the lower end of the allowable range, the optimization framework influences load demand through voltage-dependent behavior. In particular, for constant-impedance and constant-current load components, reduced voltage levels lead to lower active power consumption, reduced feeder currents, and consequently lower resistive losses, thereby improving overall system efficiency.

The CVR constraint is applied uniformly across the network and is co-optimized with power flow and topology reconfiguration decisions. This allows the model to balance the benefits of load reduction and loss minimization against voltage security considerations under varying operating conditions. Unlike conventional CVR implementations that rely on device-level control mechanisms such as discrete tap changers or heuristic voltage control rules, the proposed formulation embeds CVR directly within the optimization problem as a continuous control mechanism. This enables a systematic evaluation of CVR effectiveness under different load compositions and network configurations while preserving computational tractability.

### *2.10 Optimization Framework*

The methodology is centered on a centralized optimization framework that simultaneously determines optimal power flow variables, voltage levels, resource dispatch, and network topology. The objective is to minimize total active power losses subject to power flow equations, voltage-dependent load behavior, operational limits, and radiality constraints. Binary decision variables are introduced to represent the open or closed status of reconfigurable branches, while continuous variables capture voltages, currents, and power flows. CVR is incorporated by tightening voltage limits, allowing voltage levels to be co-optimized with topology and resource scheduling decisions.

Figure 1 illustrates multiple feasible radial configurations of the same DN obtained through topology reconfiguration. Each configuration preserves radiality while altering the status of selected tie switches, resulting in different power flow paths from the substation to downstream buses. Although the physical layout remains unchanged, alternative switching actions redistribute branch currents and voltage drops, leading to variations in active power losses. By explicitly modeling branch status using binary decision variables, the optimization framework systematically evaluates these alternative radial topologies and selects the configuration that minimizes losses while satisfying voltage and operational constraints.

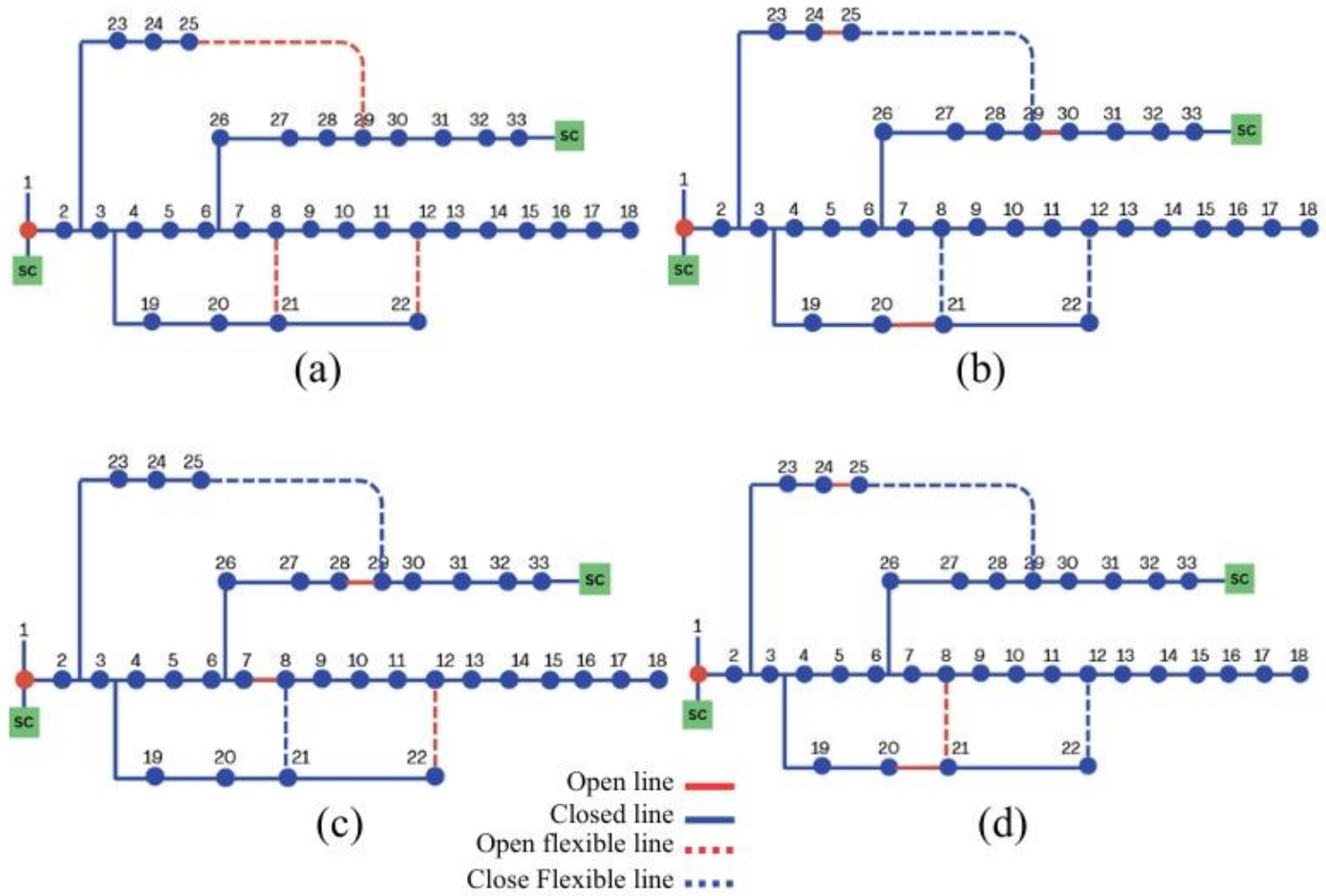

Figure 1. Feasible radial network configurations under different switching

To enable tractable solutions of the nonlinear AC power flow problem, the formulation is relaxed into a mixed-integer conic programming structure. This relaxation preserves the physical relationships between voltage, current, and power while allowing efficient solutions using commercial solvers. The resulting formulation jointly captures structural control through topology reconfiguration and operational control through CVR within a unified optimization environment.

### *2.11 Solution Strategy and Solver Implementation*

The optimization problem is implemented using the Pyomo modeling framework and solved with the Gurobi optimizer through its conic programming interface. The solver employs a branch-and-bound strategy to handle binary decision variables associated with network reconfiguration, while second-order conic constraints are used to represent nonlinear power flow relationships. This combination allows the framework to efficiently explore feasible network topologies and operating points while maintaining global feasibility with respect to voltage and current limits.

Hourly resolution is adopted to capture realistic variations in load demand and resource availability. At each time step, the optimization determines the optimal voltage profile, branch status, and resource dispatch that minimizes losses while satisfying all network constraints. Post-solution checks are performed to verify voltage compliance, current limits, and radiality of the resulting network topology.

### *2.12 Evaluation Scenarios*

To assess the effectiveness of the proposed framework, multiple operating scenarios are evaluated. These scenarios are designed to isolate the individual and combined impacts of network reconfiguration and CVR. Specifically, the following four configurations are examined:

i. SDN: a standard DN without CVR/NTR.
ii. SDNTR: a reconfigured network without CVR.
iii. CEDN: a CVR-enabled network without reconfiguration.
iv. CEDNTR: a network incorporating both CVR and NTR.

Each configuration is evaluated under different load dependency models, including constant impedance, constant current, and constant power representations, to capture varying voltage sensitivity characteristics. DERs and energy storage systems are integrated into selected scenarios to assess their interaction with voltage control and topology optimization. Performance is evaluated based on active power loss reduction, voltage profile improvement, branch loading behavior, and operational feasibility across the scheduling horizon.

### *2.13 Performance Metrics*

The primary performance metric used in this study is total active power loss over the evaluation period. Additional metrics include minimum and maximum nodal voltage magnitudes, branch current loading levels, and voltage profile uniformity along the feeder. These metrics provide a comprehensive assessment of both efficiency and operational reliability. Comparative analysis against baseline configurations is used to quantify the incremental benefits of each control strategy and to demonstrate the complementary effects of coordinated

CVR and NTR.

## 3. Case Studies

The effectiveness of the proposed optimization framework is evaluated through a series of case studies conducted on standard benchmark distribution systems. The analysis is designed to quantify the individual and combined impacts of NTR and CVR under realistic operating conditions.

### *3.1 Test system description*

The proposed framework is first evaluated using a modified IEEE 33-bus radial DN, as shown in Figure 1 [29]. This test system is widely adopted in literature due to its suitability for analyzing voltage regulation and loss minimization strategies in radial feeders. Its moderate size and well-documented characteristics make it appropriate for isolating the operational impacts of voltage control and topology optimization.

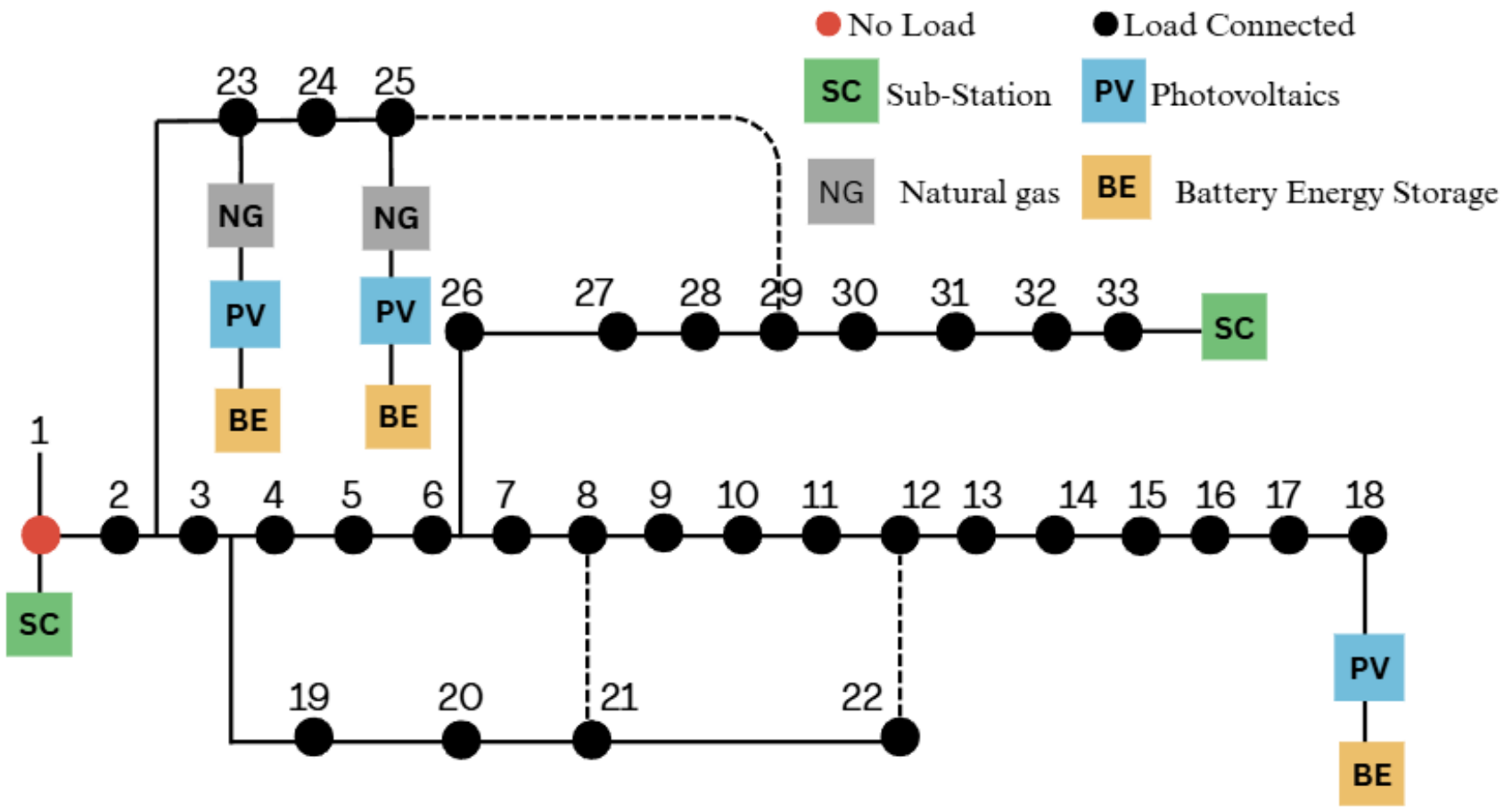

Figure 1. Modified IEEE 33-bus system.

To reflect modern distribution grid characteristics, the original network is augmented with DERs, including photovoltaic units, energy storage systems, and a dispatchable generator in addition to the main substation supply. The substation is modeled as a slack bus with fixed voltage (e.g., 1.05 p.u.), representing a strong upstream grid. This standard assumption avoids explicitly modeling the transmission network. Although PV and BESS affect feeder voltages, the substation voltage is maintained externally and remains constant. Buses are classified as load and non-load nodes, and DERs are strategically placed to provide local generation support and operational flexibility. The placement of DERs allows the study to capture both upstream and downstream voltage support effects under different loading conditions.

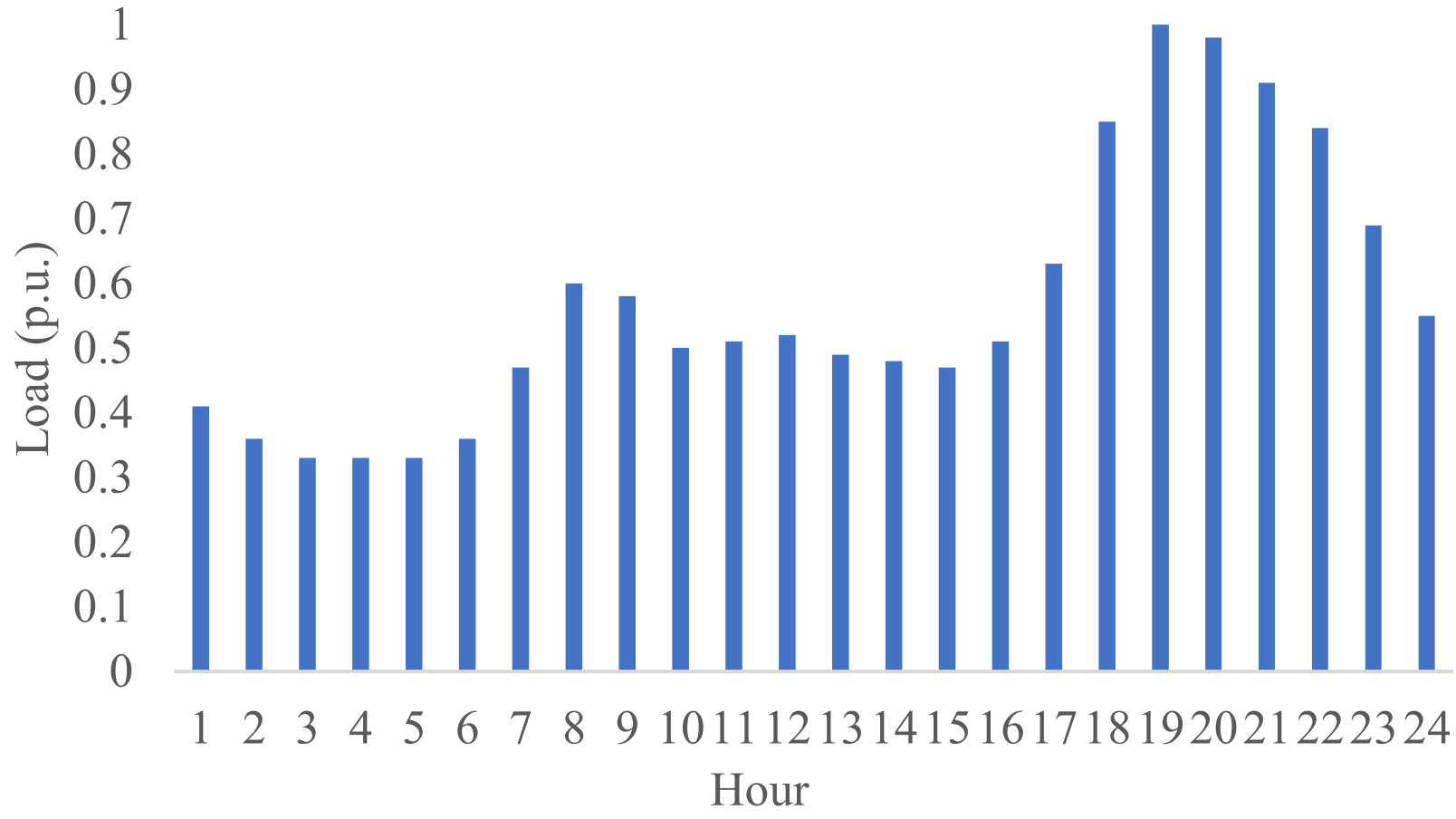

Figure 2. Hourly base load (p.u.)

The system is analyzed under multiple operating configurations to isolate the effects of structural optimization through network reconfiguration and operational optimization through CVR. Hourly load variations, shown in Figure 2, are applied to emulate realistic daily demand fluctuations. Hourly resolution is adopted to capture realistic variations in load demand and resource availability.

### *3.2 Base Case: Standard Distribution Network (SDN)*

The base case corresponds to the standard distribution network without NTR or CVR. Figure 3 presents the voltage profiles along the feeder under constant impedance (CI), constant current (CC), and constant power (CP) load models. In the absence of DERs, voltage magnitudes progressively decline from the substation toward downstream buses due to cumulative line impedance, with the minimum voltages observed near the end of the feeder. This behavior establishes a reference operating condition against which the benefits of subsequent control strategies can be quantitatively assessed. When DERs are integrated, localized power injections reduce the current drawn from the substation, resulting in improved voltage profiles across all load models. The voltage distribution becomes noticeably flatter, and minimum voltage levels increase, particularly at downstream buses. This improvement highlights the ability of DERs to partially decouple downstream voltage performance from upstream feeder impedance.

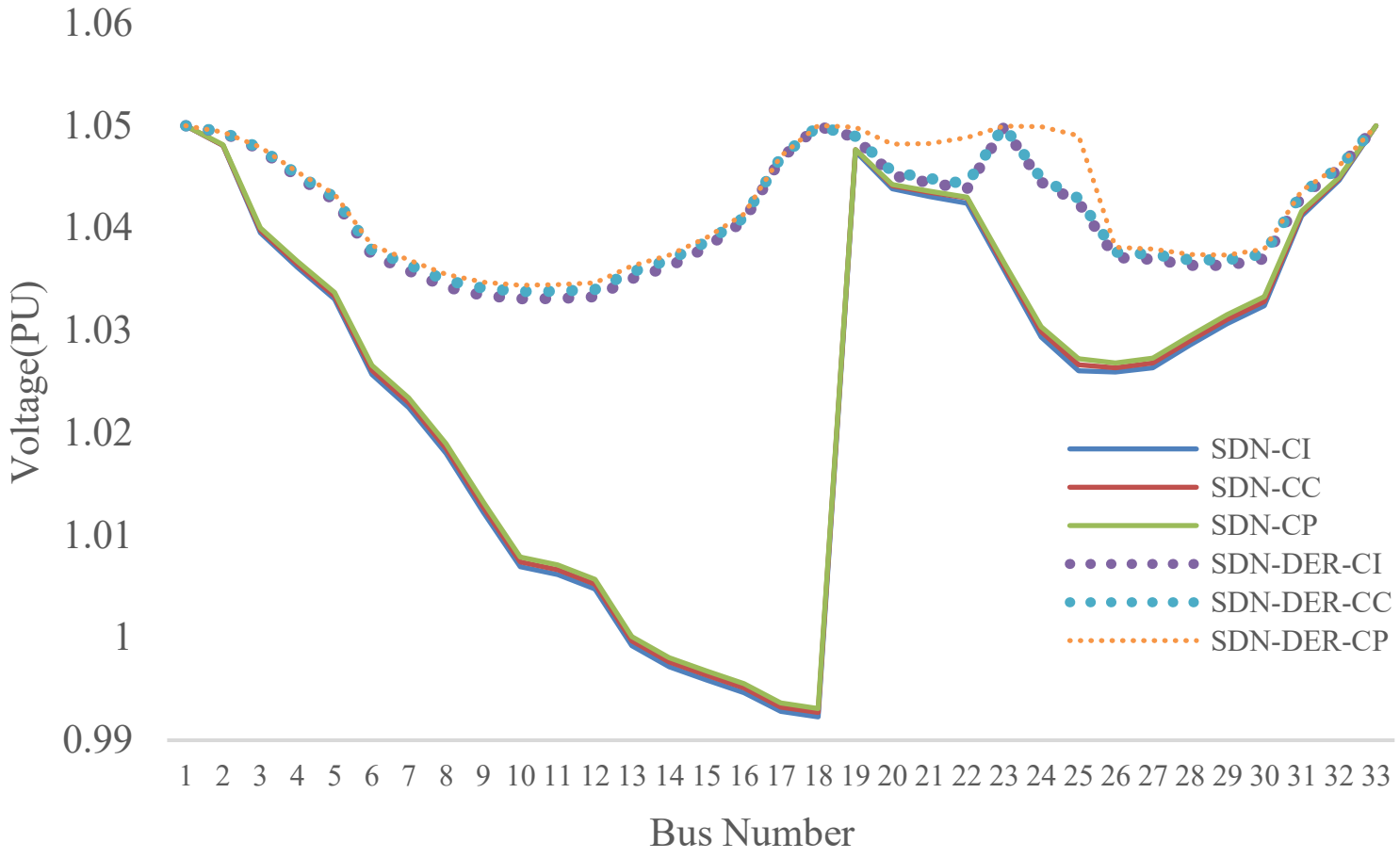

Figure 3. Voltage profile of SDN.

These results validate the correctness of the proposed optimization model and establish a reliable baseline for subsequent comparisons.

Figure 4 illustrates the corresponding hourly active power losses for the SDN case. Losses follow a typical daily trend, increasing during periods of higher demand and peaking during evening hours. With DER integration, losses are substantially reduced across the entire day, especially during peak demand periods, confirming that local generation effectively alleviates feeder loading and reduces resistive losses.

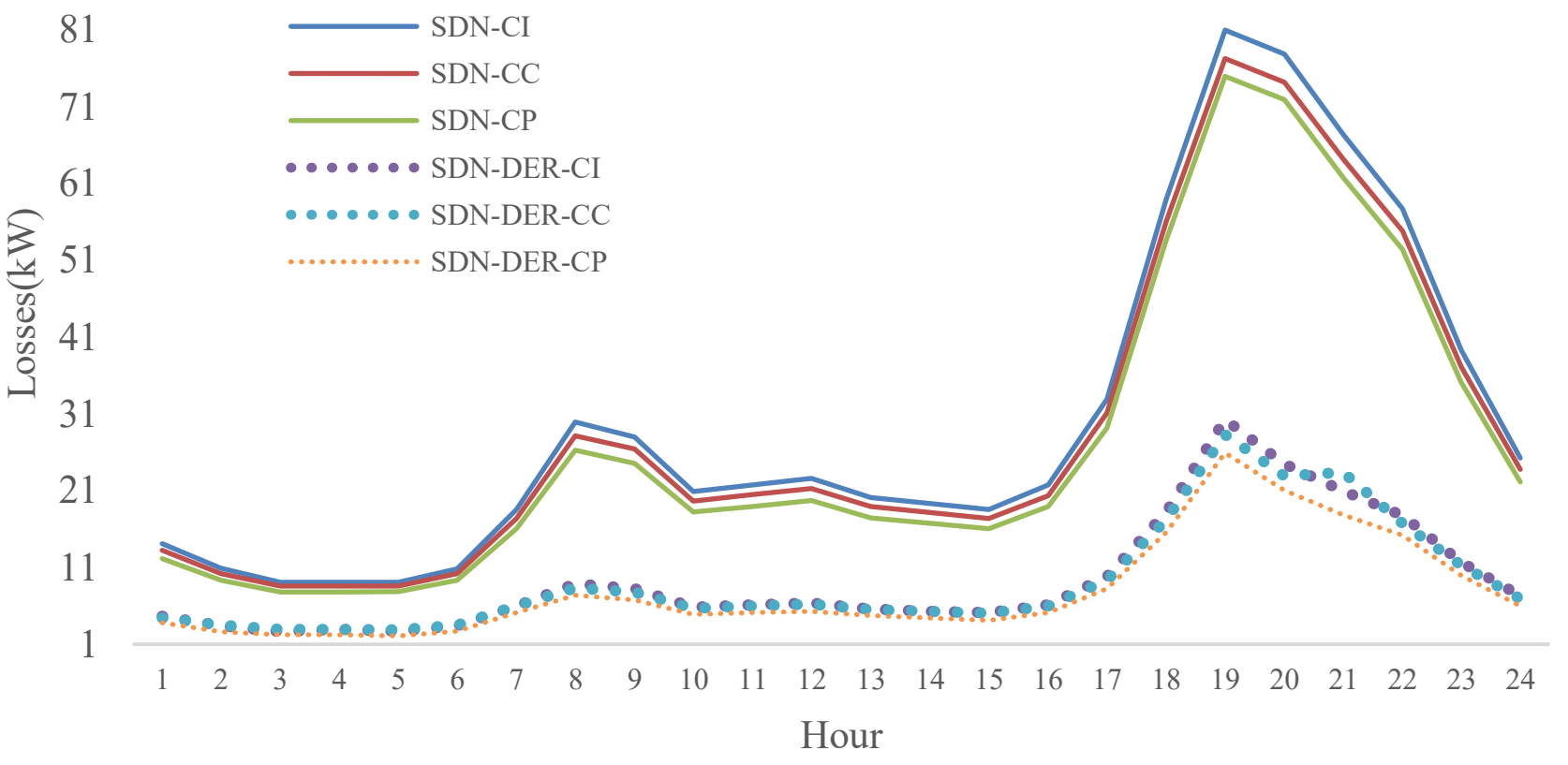

Figure 4. Hourly loss of SDN.

*3.3 Effect of Network Reconfiguration (SDN vs SDNTR)*

The impact of NTR is evaluated by comparing the SDN and SDNTR configurations. Figure 5 shows the voltage profiles for both cases under different load models, with and without DER integration. In the SDN configuration, voltage drops steadily along the feeder due to increasing electrical distance from the substation. After reconfiguration, voltage magnitudes improve consistently across the network. Notably, this improvement is achieved without modifying load demand characteristics, indicating that reconfiguration operates purely through structural optimization. This improvement is achieved by opening selected branches and redirecting power flows through lower-impedance paths, thereby reducing current loading on critical sections.

When DERs are present, voltage profiles further improve, and the combined effect of reconfiguration and local generation results in enhanced voltage stability across all load representations. These results demonstrate that topology reconfiguration and DER integration provide complementary voltage support mechanisms.

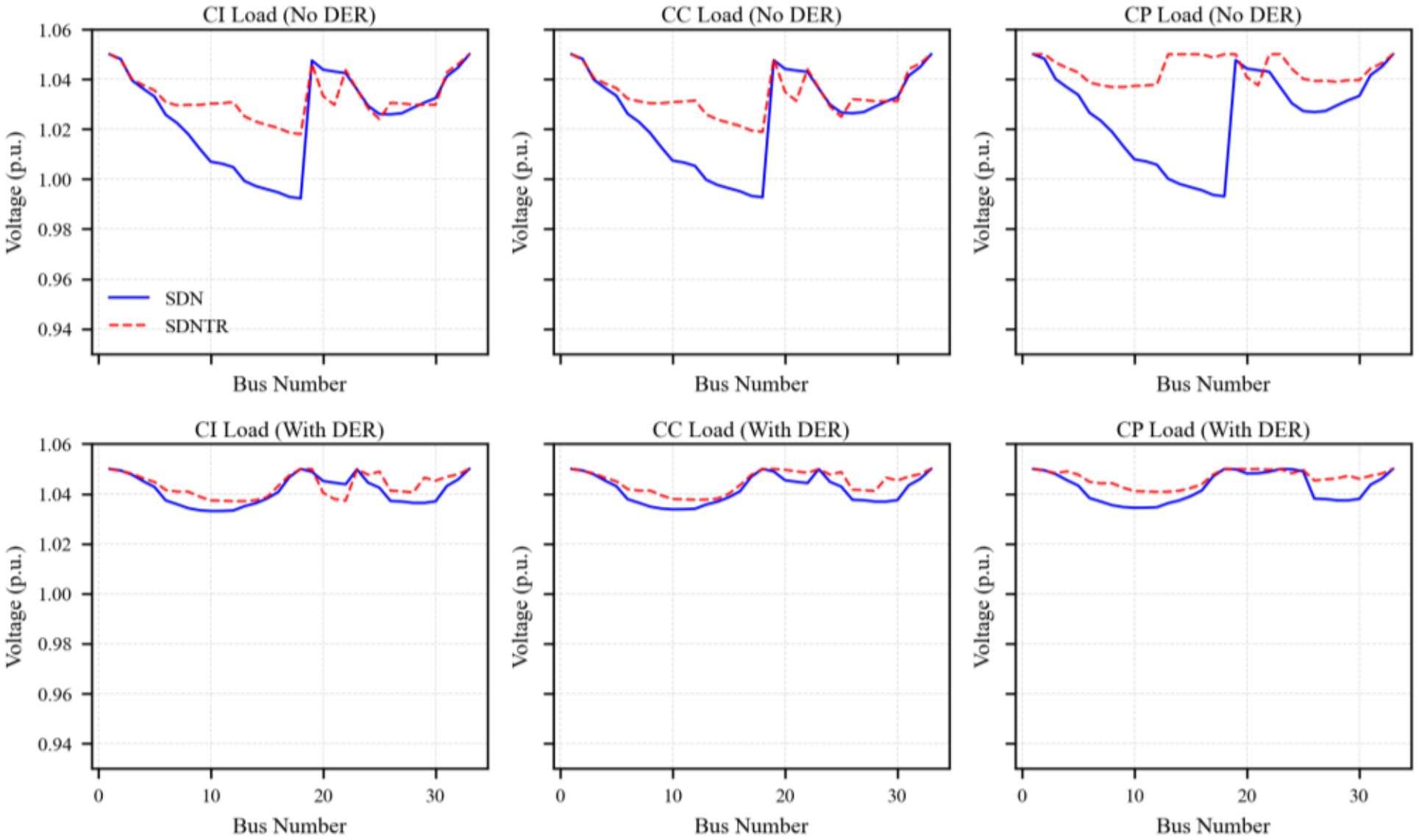

Figure 5. Voltage profiles for SDN vs SDNTR, with and without DER.

Table I summarizes the corresponding loss reductions achieved through network reconfiguration. Without DER integration, reconfiguration yields an average loss reduction of approximately 7%. When DERs are present, the reduction increases to approximately 10%. Importantly, the effectiveness of reconfiguration remains consistent across all load models, indicating that structural optimization provides a robust loss minimization mechanism independent of load voltage sensitivity.

TABLE I. COMPARISON OF ACTIVE POWER LOSS REDUCTION DUE TO NETWORK RECONFIGURATION (SDN → SDNTR) UNDER DIFFERENT LOAD MODELS WITH AND WITHOUT DER

| | ***Without DER*** | | | ***With DER*** | | |
|---|---|---|---|---|---|---|
| **Configuration** | $P_{load}$ (MW) | $P_{loss}$ (MW) | $P_{loss}$ (%) | $P_{load}$ (MW) | $P_{loss}$ (MW) | $P_{loss}$ (%) |
| **SDN-CI** | 54.53 | 0.73 | 1.33 | 55.60 | 0.23 | 0.40 |
| **SDNTR-CI** | 54.86 | 0.69 | 1.25 | 55.71 | 0.21 | 0.37 |
| **SDN-CC** | 52.69 | 0.69 | 1.31 | 53.20 | 0.22 | 0.41 |
| **SDNTR-CC** | 52.87 | 0.64 | 1.21 | 53.28 | 0.19 | 0.36 |
| **SDN-CP** | 50.89 | 0.65 | 1.28 | 50.89 | 0.20 | 0.39 |
| **SDNTR-CP** | 50.89 | 0.59 | 1.17 | 50.89 | 0.18 | 0.35 |

### *3.4 Effect of CVR (SDN vs CEDN)*

The impact of CVR is examined by comparing the SDN and CEDN configurations. Figure 6 presents the voltage profiles under CI, CC, and CP load models. Activation of CVR results in a controlled reduction of feeder voltage within regulatory limits, typically between 3% and 5% relative to the SDN case. This voltage reduction is spatially uniform along the feeder, reflecting the system-wide nature of the CVR constraint.

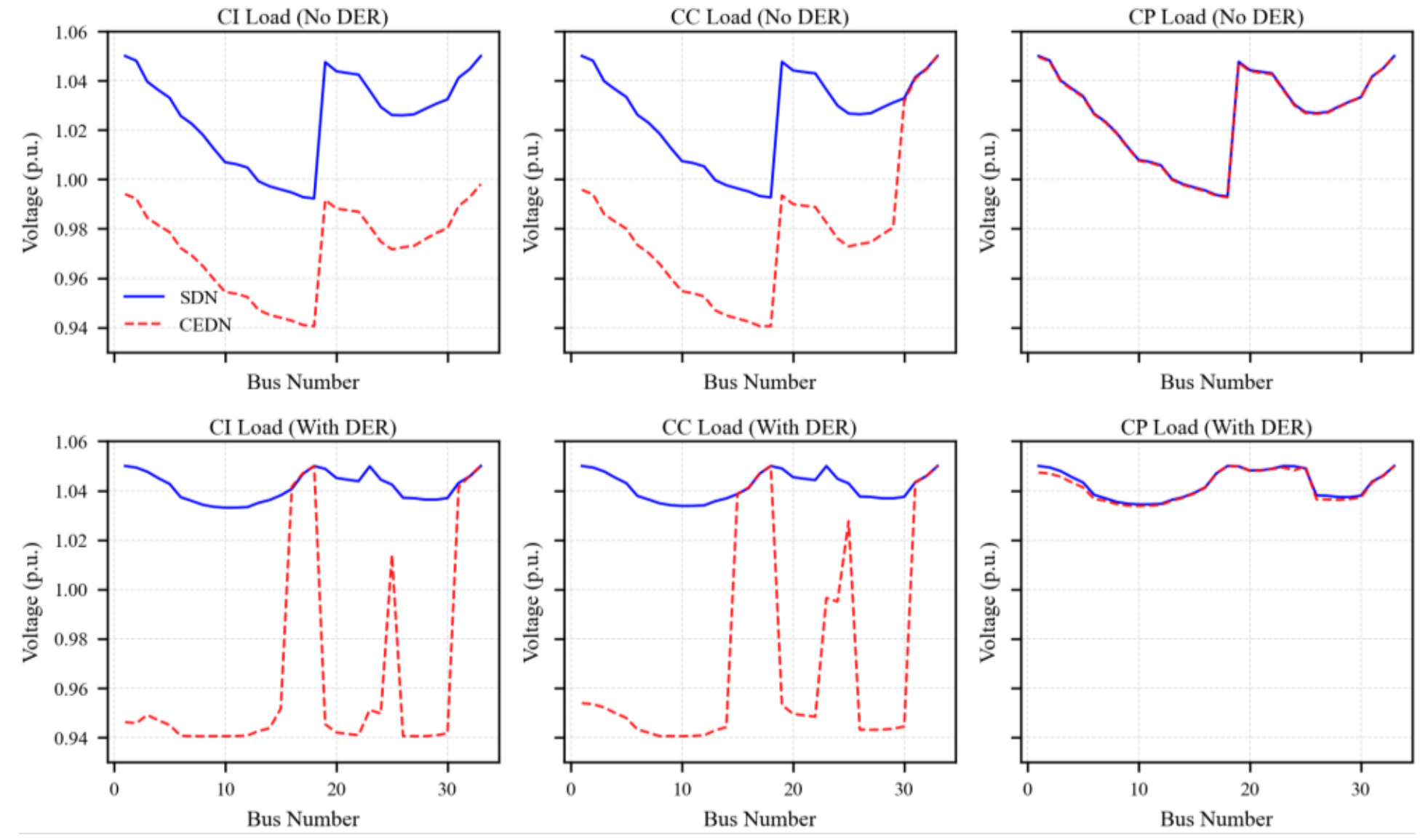

Figure 6. Voltage profiles for SDN vs CEDN, with and without DER.

The quantitative impact of CVR is summarized in Table II. Under CI loading, CVR achieves significant reductions in both load demand and active power losses, with further improvement observed when DERs are present. The Under voltage-dependent load models, particularly CI and CC, reduced voltage levels lead to lower active power demand and decreased branch currents. As a result, active power losses are reduced without violating voltage constraints. In contrast, the CP load model exhibits minimal change in both voltage profile and losses, as its demand is largely insensitive to voltage variations. This behavior reinforces the importance of accurate load modeling when evaluating CVR performance.

CC model exhibits moderate reductions, while the CP model shows negligible change. These results confirm that the effectiveness of CVR is strongly dependent on load voltage sensitivity and is enhanced by DER integration.

TABLE II. COMPARISON OF ACTIVE POWER LOSS REDUCTION DUE TO NETWORK RECONFIGURATION (SDN → SDNTR) UNDER DIFFERENT LOAD MODELS WITH AND WITHOUT DER

| | ***Without DER*** | | ***With DER*** | |
|---|---|---|---|---|
| **Load Type** | Load Reduction (%) | Loss Reduction (%) | Load Reduction (%) | Loss Reduction (%) |
| **Constant Impedance** | 13.70 | 15.07 | 14.23 | 21.74 |
| **Constant Current** | 4.12 | 1.45 | 6.47 | 9.09 |

### *3.5 Combined Impact of CVR and Reconfiguration (CEDNTR)*

The combined application of CVR and NTR is evaluated under the CEDNTR configuration. Figure 7 compares voltage profiles across SDN, SDNTR, CEDN, and CEDNTR cases. While SDNTR improves voltage levels through structural optimization and CEDN reduces voltage in a controlled manner to lower demand, the CEDNTR configuration achieves the most uniform voltage profile across the feeder. This uniformity indicates that voltage moderation and current redistribution act synergistically rather than competitively. Figure 8 presents the corresponding hourly active power losses.

The CEDNTR configuration consistently yields the lowest loss values across all load models and operating conditions. The complementary interaction between CVR and reconfiguration enables both demand reduction and efficient current redistribution, resulting in superior loss minimization compared to either strategy applied independently. These results highlight the value of coordinated structural and operational control in DNs.

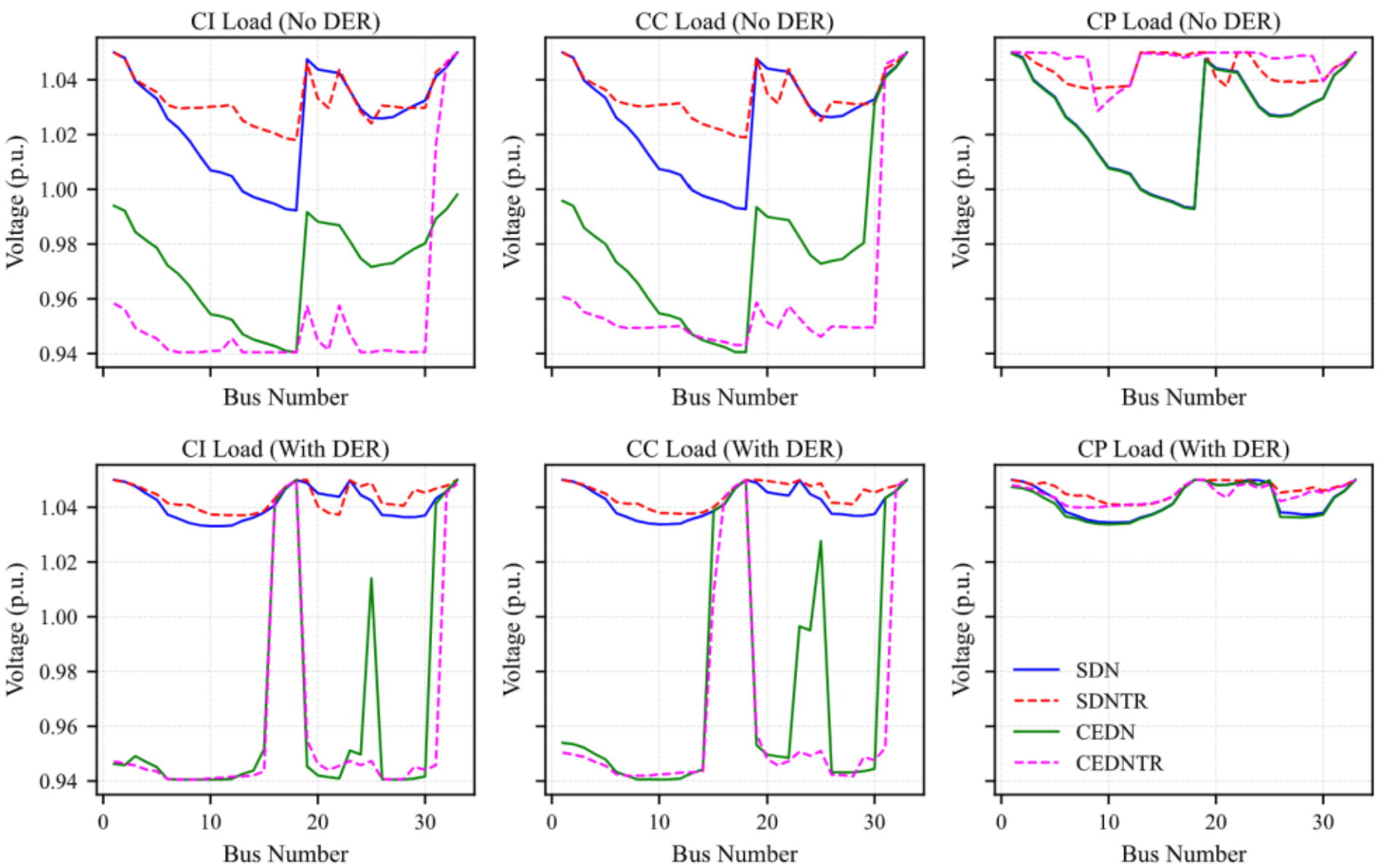

Figure 7. Voltage profiles for SDN, SDNTR, CEDN, CEDNTR (with and without DER) against different load types.

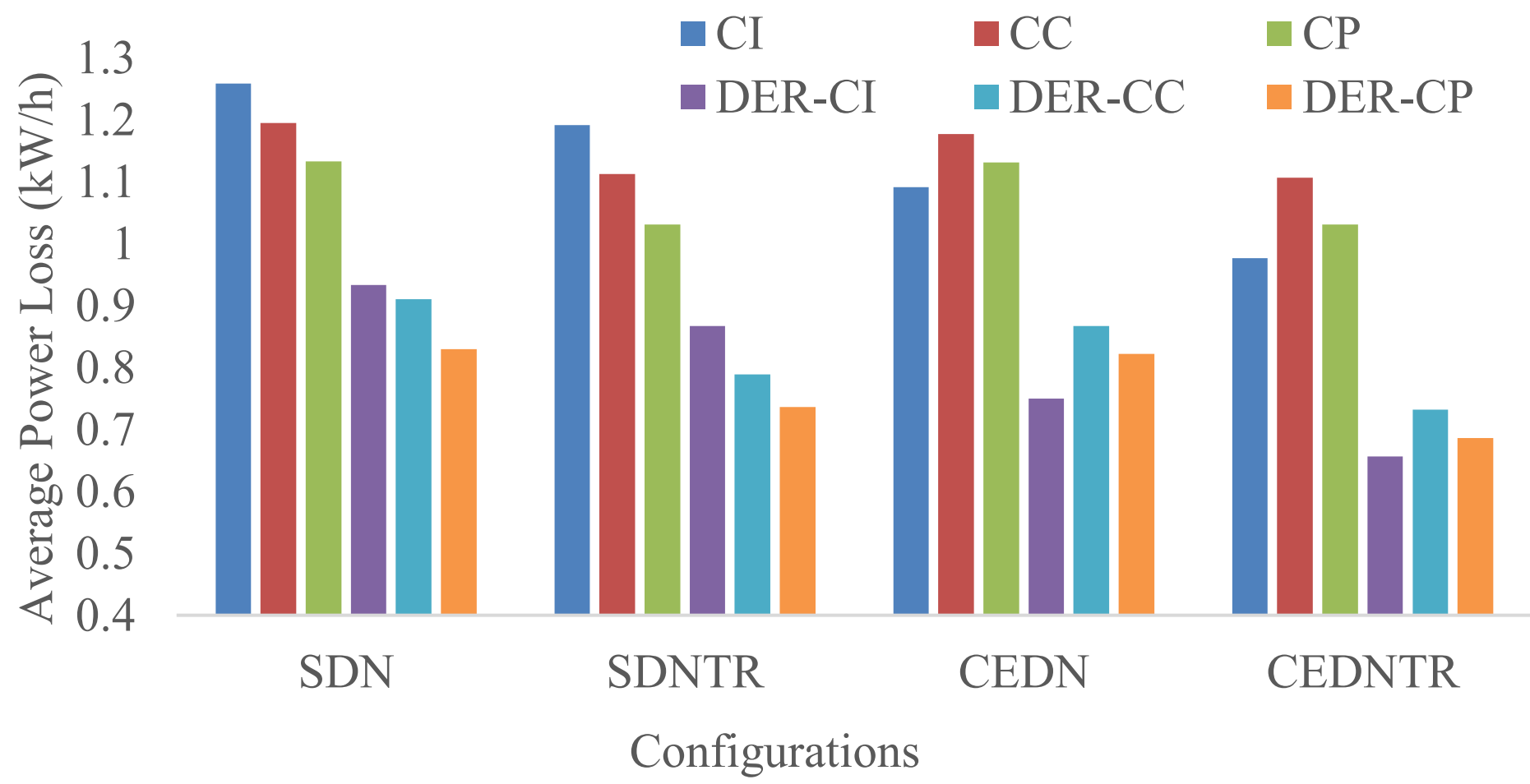

Figure 8. Hourly loss comparison: SDN, SDNTR, CEDN, CEDNTR (with and without DER).

Table III lists the optimal switching combinations identified under SDNTR and CEDNTR configurations. The selected switching patterns are largely consistent, indicating that reconfiguration decisions are primarily driven by feeder structure, with CVR exerting a secondary influence when DERs are present.

TABLE III. LINES OPEN DURING OPTIMAL NETWORK RECONFIGURATION FOR EACH CONFIGURATION

| Configuration | CI Load | CC Load | CP Load | DER-CI | DER-CC | DER-CP |
|---|---|---|---|---|---|---|
| **SDNTR** | 8, 28, 29, 33 | 7, 8, 28, 29 | 8, 28, 29, 33 | 7, 8, 12, 28 | 7, 12, 28, 34 | 7, 12, 28, 34 |
| **CEDNTR** | 8, 28, 29, 33 | 8, 28, 29, 33 | 7, 8, 28, 29 | 7, 8, 12, 28 | 7, 8, 12, 28 | 7, 8, 12, 28 |

## 4. Robustness and Scalability Evaluation of the Proposed Framework

To evaluate the scalability and robustness of the proposed optimization framework under realistic network complexity, the IEEE 123-bus distribution feeder is selected as a large-scale test system. Although the original feeder is inherently unbalanced, it is modeled as a balanced three-phase network to maintain computational tractability while preserving feeder depth, connectivity, and operational diversity. Voltage-dependent ZIP load modeling is retained to ensure consistency with the earlier case studies and to accurately reflect demand sensitivity under voltage regulation.

As shown in Figure 10, the modified IEEE 123-bus feeder [30] consists of a deep radial structure with multiple laterals and a combination of load and non-load buses. The system includes multiple distributed resources and storage units placed across different feeder locations. These installations are assumed to follow realistic deployment patterns, where DER integration is largely driven by end-user adoption and is not centrally coordinated, resulting in a dispersed and non-uniform spatial distribution across the network. In addition, the feeder includes a large number of sectionalizing and tie switches that provide extensive reconfiguration flexibility. These characteristics significantly increase the dimensionality and combinatorial complexity of the optimization problem compared to smaller benchmark feeders.

Relative to the IEEE 33-bus system, the IEEE 123-bus feeder introduces a more challenging operational environment due to longer electrical distances, increased switching options, and a wider range of feasible power flow paths. This makes it well suited for assessing whether the proposed framework maintains its performance benefits as network size and complexity increase. Results obtained on this system demonstrate that the coordinated application of voltage regulation and topology optimization continues to deliver substantial performance gains, confirming the robustness and scalability of the proposed approach.

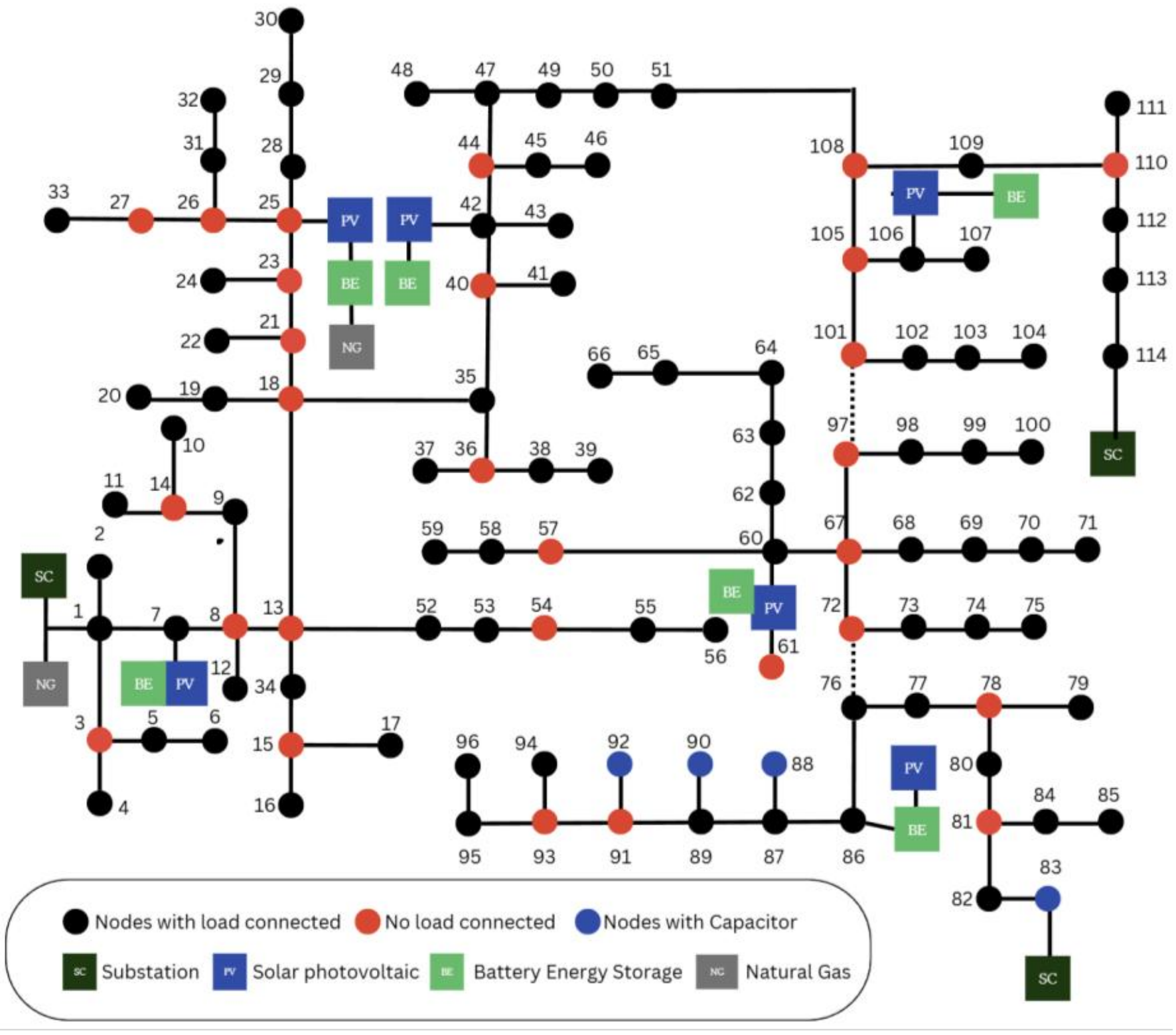

Figure 9. IEEE modified 123 bus system.

*4.1 Active power loss minimization*

Figure 11 compares total active power losses for SDN, SDNTR, CEDN, and CEDNTR configurations on the IEEE 123-bus feeder. In the SDN configuration, total feeder losses are approximately 17 kW for a system load of 3.14 MW. Topology reconfiguration alone reduces losses to approximately 16 kW, corresponding to a 5.9% reduction relative to the base case. Voltage regulation applied independently yields a larger reduction, lowering losses to approximately 14.8 kW and achieving a 12.9% improvement. When both strategies are applied simultaneously, losses are reduced to approximately 13.5 kW, representing a 20.6% reduction relative to the SDN case.

The observed performance improvement under the combined strategy is attributed to the complementary interaction between CVR and NTR. Specifically, CVR reduces voltage-dependent load demand, leading to lower feeder currents and alleviating line loading across the network. This reduction in current expands the feasible solution space for topology reconfiguration, allowing switching actions that further redistribute power flows along lower-resistance paths. Conversely, NTR improves the voltage profile by mitigating voltage drops across heavily loaded branches, enabling a larger portion of the network to operate closer to the lower voltage bound. This enhances the effectiveness of CVR by increasing the proportion of voltage-sensitive load that can be influenced.

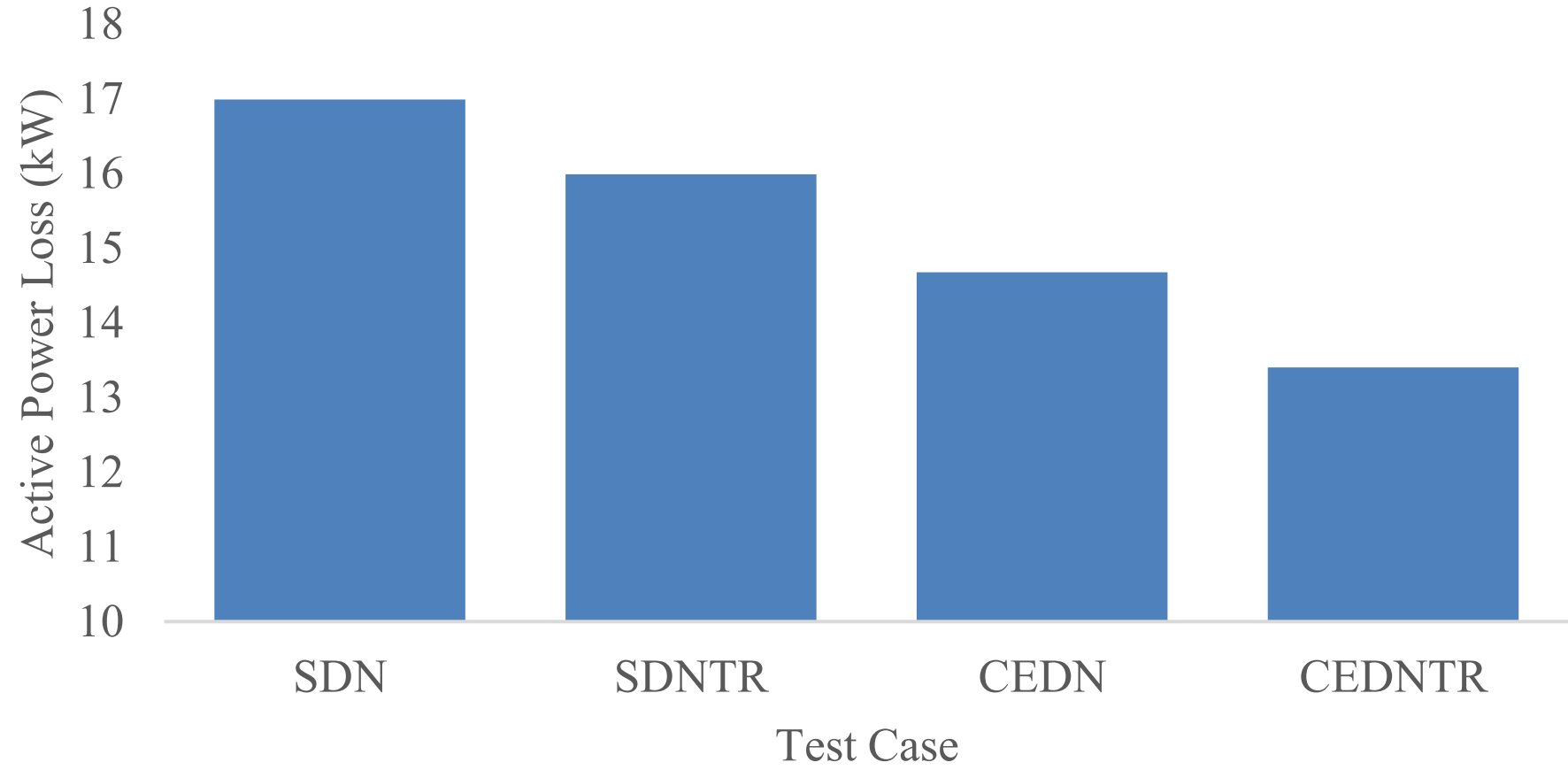

Figure 10. Active power loss.

As a result, the coordinated application of CVR and NTR produces a synergistic effect, yielding greater loss reduction than the sum of individual contributions. These results confirm that structural (NTR) and operational (CVR) strategies are inherently complementary, and that their joint optimization leads to improved system performance, even in large-scale distribution networks.

*4.2 Branch Loading Heatmap of the IEEE 123-Bus Feeder*

Figure 12 illustrates the line loading distribution for the SDN configuration, where the network operates under a fixed radial topology. In this case, power flows are constrained along predefined feeder paths, resulting in pronounced loading concentration along primary backbone branches supplying downstream laterals. Several lines exhibit relatively high utilization levels, indicating localized congestion and limited flexibility in redistributing power flows.

Figure 13 presents the corresponding results for the CEDNTR configuration, where network reconfiguration and voltage regulation are jointly applied. The introduction of flexible switching enables the formation of alternative power flow paths, as indicated by the activated tie-lines, allowing load to be redistributed away from heavily utilized branches. This results in a more uniform loading profile across the network.

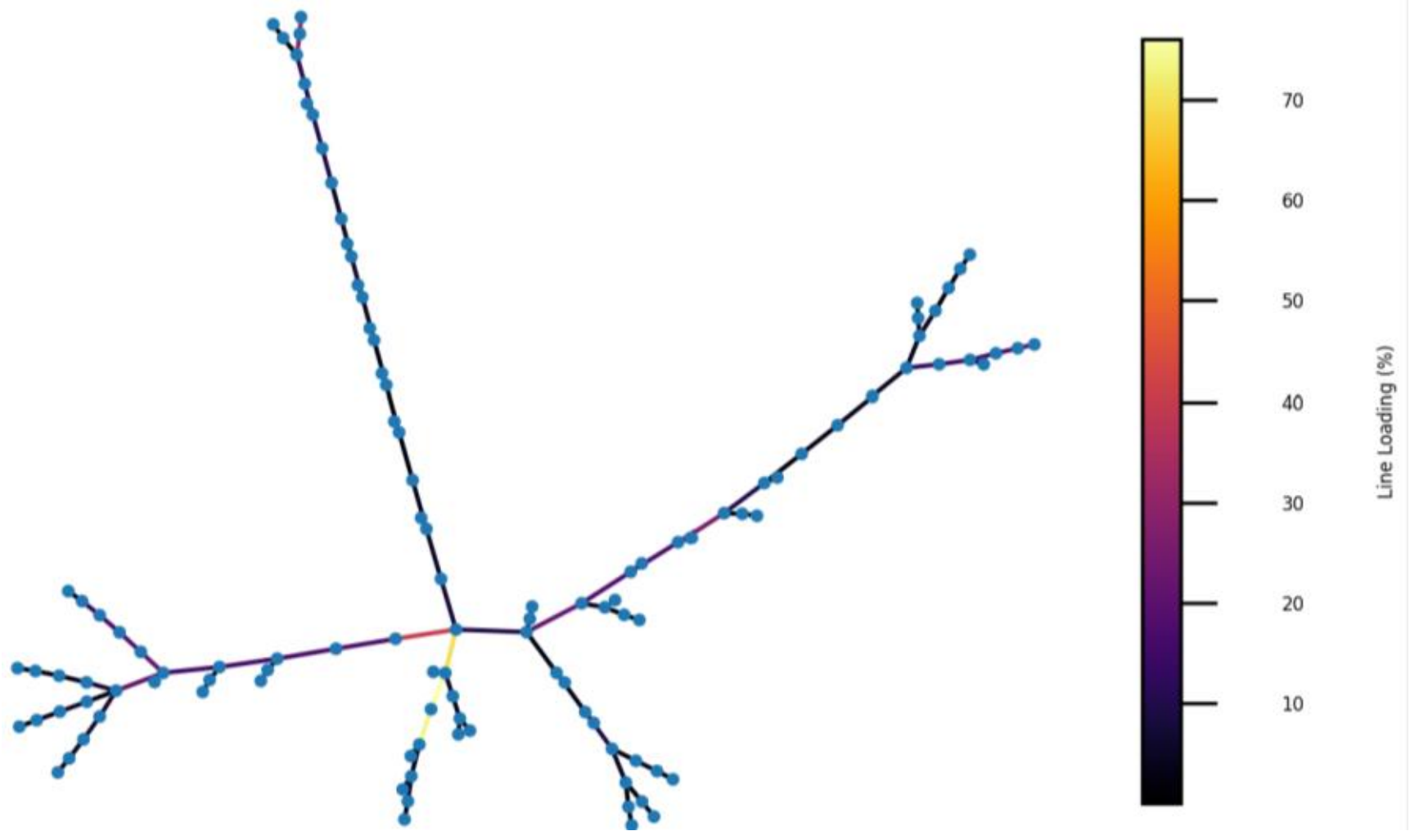

Figure 11. SDN line loading.

In addition to structural redistribution, CVR contributes to an overall reduction in feeder current by lowering voltage-dependent load demand. As a result, both the magnitude and distribution of line loading are improved. The maximum line loading decreases from approximately 75% in the SDN case to about 62% in the CEDNTR case, corresponding to a 17% reduction. Furthermore, the number of heavily loaded lines is visibly reduced, indicating improved utilization of network capacity.

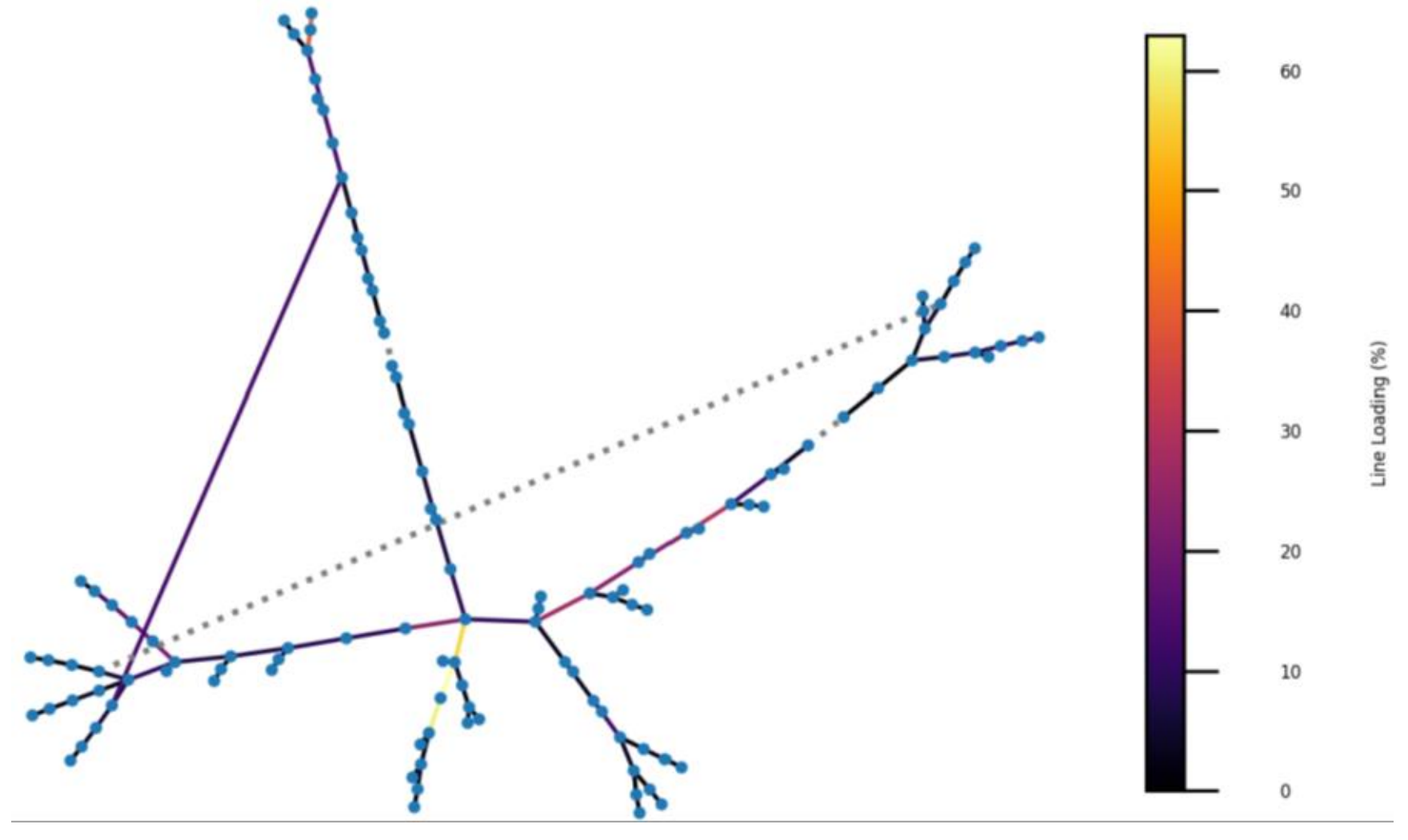

Figure 12. CEDNTR line loading.

This reduction in line loading directly supports the observed loss reduction trends presented in Figure 11. Lower current magnitudes lead to reduced $I^2R$ losses, while improved flow distribution minimizes localized congestion and associated losses. Thus, the combined CVR-NTR strategy simultaneously reduces both the magnitude and concentration of current flows across the network. Overall, these results confirm that CVR and NTR exhibit a synergistic interaction: CVR reduces current levels, thereby expanding the feasible space for reconfiguration, while NTR redistributes power flows to improve voltage profiles and enable more effective voltage reduction. This coordinated behavior enhances thermal margins, reduces losses, and improves operational robustness in large-scale distribution systems.

## 5. Conclusion

This paper presented a coordinated optimization framework for the day-ahead operational planning of radial distribution networks by integrating CVR and NTR. By capturing the inherent interaction between voltage control and network topology within a unified mixed-integer conic programming formulation, the proposed approach enables simultaneous adjustment of voltage levels and feeder configuration under realistic operational constraints. The results highlight that CVR-driven load reduction and NTR-based line loading redistribution are strongly interdependent, and their coordinated application leads to more effective system operation than independent implementations.

The effectiveness of the proposed framework is demonstrated on the IEEE 33-bus distribution system, where the coordinated approach achieves the lowest active power losses compared to individual strategies. In addition, results on the IEEE 123-bus system show that losses are reduced from approximately 17 kW in the base case to about 13.4 kW under the coordinated CVR-NTR implementation, outperforming both standalone voltage control and topology reconfiguration. The results confirm that neither CVR nor NTR alone can achieve the same level of performance as their coordinated application. Furthermore, the IEEE 123-bus system is used to assess the robustness and scalability of the approach, confirming its effectiveness under increased network size and complexity. Overall, the proposed framework provides a practical and scalable solution for improving distribution system efficiency through integrated, system-level operational planning.

## References

[1] J. Marcos, L. Marroyo, E. Lorenzo, D. Alvira, and E. Izco, “Power output fluctuations in large scale PV plants: One year observations with one second resolution and a derived analytic model,” Progress Photovoltaics: Res. Appl., vol. 19, no. 2, pp. 218–227, 2011.

[2] B. Palmintier et al., “On the path to sunshot: Emerging issues and challenges in integrating solar with the distribution system,” Nat. Renew. Energy Lab. [Online]. Available: http://www.nrel.gov/docs/fy16osti/65331. pdf

[3] Y. P. Agalgaonkar, B. C. Pal, and R. A. Jabr, “Distribution voltage control considering the impact of PV generation on tap changers and autonomous regulators,” IEEE Trans. Power Syst., vol. 29, no. 1, pp. 182–192, Jan. 2014.

[4] R. Yan and T. K. Saha, “Investigation of voltage stability for residential customers due to high photovoltaic penetrations,” IEEE Trans. Power Syst., vol. 27, no. 2, pp. 651–662, May 2012.

[5] R. Yan, B. Marais, and T. K. Saha, “Impacts of residential photovoltaic power fluctuation on on-load tap changer operation and a solution using DSTATCOM,” Electric Power Syst. Res., vol. 111, pp. 185–193, 2014. [Online]. Available: http://dx.doi.org/10.1016/j.epsr.2014.02.020

[6] M. Kraiczy, A. L. Fakhri, T. Stetz, and M. Braun, “Do it locally: Local voltage support by distributed generation-A management summary,” Int. Energy Agency, Paris, France, Tech. Rep. IEA-PVPS T14-08, vol. 2017, 2017.

[7] A. Kulmala, S. Repo, and P. Jarventausta, “Coordinated voltage control in distribution networks including several distributed energy resources,” IEEE Trans. Smart Grid, vol. 5, no. 4, pp. 2010–2020, Jul. 2014.

[8] R. Tonkoski, L. A. Lopes, and T. H. El-Fouly, “Coordinated active power curtailment of grid connected PV inverters for overvoltage prevention,” IEEE Trans. Sustain. Energy, vol. 2, no. 2, pp. 139–147, Apr. 2011.

[9] V. Calderaro, V. Galdi, F. Lamberti, and A. Piccolo, “A smart strategy for voltage control ancillary service in distribution networks,” IEEE Trans. Power Syst., vol. 30, no. 1, pp. 494–502, Jan. 2015.

[10] A. B. Eltantawy and M. M. Salama, “Management scheme for increasing the connectivity of small-scale renewable DG,” IEEE Trans. Sustain. Energy, vol. 5, no. 4, pp. 1108–1115, Oct. 2014.

[11] M. E. Baran and F. F.Wu, “Network reconfiguration in distribution systems for loss reduction and load balancing,” IEEE Trans. Power Del., vol. 4, no. 2, pp. 1401–1407, Apr. 1989.

[12] R. A. Jabr, R. Singh, and B. C. Pal, “Minimum loss network reconfiguration using mixed-integer convex programming,” IEEE Trans. Power Syst., vol. 27, no. 2, pp. 1106–1115, May 2012.

[13] J. A. Taylor and F. S. Hover, “Convex models of distribution system reconfiguration,” IEEE Trans. Power Syst., vol. 27, no. 3, pp. 1407–1413, Aug. 2012.

[14] A. Ahuja, S. Das, and A. Pahwa, “An AIS-ACO hybrid approach for multi-objective distribution system reconfiguration,” IEEE Trans. Power Syst., vol. 22, no. 3, pp. 1101–1111, Aug. 2007.

[15] A. Asrari, S. Lotfifard, and M. Ansari, “Reconfiguration of smart distribution systems with time varying loads using parallel computing,” IEEE Trans. Smart Grid, vol. 7, no. 6, pp. 2713–2723, Nov. 2016.

[16] A. Kavousi-Fard and T. Niknam, “Optimal distribution feeder reconfiguration for reliability improvement considering uncertainty,” IEEE Trans. Power Del., vol. 29, no. 3, pp. 1344–1353, Jun. 2014

[17] Z. Tian, W. Wu, B. Zhang, and A. Bose, “Mixed-integer second-order cone programing model for VAR optimisation and network reconfiguration in active distribution networks,” IET Gener. Transm. Distrib., vol. 10, no. 8, pp. 1938–1946, May 2016.

[18] K. Turitsyn, P. Sulc, S. Backhaus, and M. Chertkov, “Options for control of reactive power by distributed photovoltaic generators,” Proc. IEEE, vol. 99, no. 6, pp. 1063–1073, Jun. 2011.

[19] H. Yuan, F. Li, Y. Wei, and J. Zhu, “Novel linearized power flow and linearized OPF models for active distribution networks with application in distribution LMP,” IEEE Trans. Smart Grid, vol. 9, no. 1, pp. 438–448, Jan. 2016

[20] Rida Fatima, Hassan Zahid Butt, and Xingpeng Li, “Optimal Dynamic Reconfiguration of Distribution Networks”, 55th North American Power Symposium, Asheville, NC, USA, Oct. 2023.

[21] Arun Venkatesh Ramesh and Xingpeng Li, “Network Reconfiguration Impact on Renewable Energy System and Energy Storage System in Day-Ahead Scheduling,” IEEE PES General Meeting 2021, (Virtually), Washington, DC, USA, Jul. 2021.

[22] American National Standard for Electric Power Systems and Equipment Voltage Ratings (60 Hertz), Amer. Nat. Stand. Inst., New York, NY, USA, 2016.

[23] Z. Wang and J. Wang, “Review on implementation and assessment of conservation voltage reduction,” IEEE Trans. Power Syst., vol. 29, no. 3, pp. 1306–1315, May 2014.

[24] D. Kirshner, “Implementation of conservation voltage reduction at commonwealth edison,” IEEE Trans. Power Syst., vol. 5, no. 4, pp. 1178–1182, Nov. 1990.

[25] Z. Wang, M. Begovic, and J. Wang, “Analysis of conservation voltage reduction effects based on multistage SVR and stochastic process,” IEEE Trans. Smart Grid, vol. 5, no. 1, pp. 431–439, Jan. 2014.

[26] J. Wang, C. Chen, and X. Lu, “Guidelines for implementing advanced distribution management systems-requirements for DMS integration with DERMS and microgrids,” Argonne Nat. Lab., Argonne, IL, USA, Rep. ANL/ESO-15/15, 2015.

[27] Q. Shi, W. Feng, Q. Zhang, X. Wang, and F. Li, “Overvoltage mitigation through volt-var control of distributed PV systems,” in Proc. IEEE Power Energy Soc. Tranm. Distrib. (T&D), 2020, pp. 1–5.

[28] F. Ding et al., “Photovoltaic impact assessment of smart inverter volt-var control on distribution system conservation voltage reduction and power

quality," Nat. Renew. Energy Lab., Golden, CO, USA, Rep. NREL/TP5D00-67296, 2016.

[29] Rida Fatima, Linhan Fang, and Xingpeng Li, "A Reliability-Cost Optimization Framework for EV and DER Integration in Standard and Reconfigurable Distribution Network Topologies", arXiv, Nov. 2025.

[30] K. P. Schneider, B. A. Mather, B. C. Pal, C. W. Ten, G. J. Shirek, H. Zhu, J. C. Fuller, J. L. R. Pereira, L. F. Ochoa, L. R. de Araujo, R. C. Dugan, S. Matthias, S. Paudyal, T. E. McDermott, and W Kersting, "Analytic Considerations and Design Basis for the IEEE Distribution Test Feeders," IEEE Transactions on Power Systems, vol. PP, no. 99, pp. 1-1, 2017.